\documentclass[reprint]{aastex62}

\slugcomment{(Accepted for publication in the Astronomical Journal)}
\shorttitle{PHOTOMETRIC TRANSIT FOLLOW UP OBSERVATIONS}
\shortauthors{CHAKRABARTY \& SENGUPTA}

\usepackage{natbib}
\begin{document}

\title{PRECISE PHOTOMETRIC TRANSIT FOLLOW-UP OBSERVATIONS OF FIVE CLOSE-IN 
EXOPLANETS : UPDATE ON THEIR PHYSICAL PROPERTIES}

\correspondingauthor{Aritra Chakrabarty}
\email{$^\dagger$aritra@iiap.res.in}

\author[0000-0001-6703-0798]{Aritra Chakrabarty$^\dagger$}
\affil{Indian Institute of Astrophysics, Koramangala 2nd Block,
Bangalore 560034, India}
\affil{University of Calcutta, Salt Lake City, JD-2,
Kolkata 750098, India}
%\email{$^*$aritra@iiap.res.in}

\author{Sujan Sengupta$^\ddagger$}
\affil{Indian Institute of Astrophysics, Koramangala 2nd Block,
Bangalore 560034, India}
\email{$^\ddagger$sujan@iiap.res.in}

\vspace{0.05\textwidth}
\begin{abstract}
We report the results of the high precision photometric follow-up observations
of five transiting hot jupiters - WASP-33b, WASP-50b,  WASP-12b, HATS-18b
and HAT-P-36b. The observations are made from the 2m Himalayan Chandra 
Telescope at Indian Astronomical Observatory, Hanle and the 1.3m J. C.
Bhattacharyya Telescope at Vainu Bappu Observatory, Kavalur. This exercise is
a part of the capability testing of the two telescopes and their back-end
instruments. Leveraging the large aperture of both the telescopes used, the
images taken during several nights were used to produce 
the transit light curves with high photometric S/N ($>200$) by performing
differential photometry. In order to reduce the fluctuations in the transit light curves
 due to various sources such as stellar activity, varying sky transparency etc. we preprocessed them using wavelet denoising and applied Gaussian process correlated noise modeling technique while modeling the transit light curves. To demonstrate the efficiency of the wavelet denoising process we have also included the results without the denoising process. A state-of-the-art algorithm used for modeling the transit light curves provided the physical parameters of the planets with more precise values than reported earlier.
\end{abstract}

%\keywords{stars: low-mass --- brown dwarfs --- stars: atmospheres} 

\section{Introduction}
Since the discovery of the first transiting exoplanet HD 209458b
\citep{charbonneau00,henry00} it was realized that transit photometric
observations are necessary to obtain a number of essential information 
that cannot be otherwise derived from the other methods of detection.
The precise understanding on the radius, surface gravity, orbital distance,
etc. helps us comprehend their formation and evolution. 

Proper characterization of the already discovered exoplanets calls for
repeated follow-up observations during both in-transit and out-transit epoch. 
In this regard the ground-based telescopes are appropriate especially in case of
a transit by a close-in (hence short period) gas giant planet around a main
sequence star. However, in order to observe a transient event like a transit,
a coordinated set of observations around the globe will prove to be highly
effective by ensuring the coverage of a transit event regardless of its 
ephemeris of occurrence. In this regard, the astronomical facilities of
India such as Indian Astronomical Observatory (IAO, $78^\circ$ $57'$ E, 
$32^\circ$ $46'$ N), Vainu Bappu Observatory 
(VBO, $78^\circ$ $50'$ E, $12^\circ$ $34'$ N) can fill in the missing 
longitudinal coverage. We, therefore, observed 
the transit events of a few hot Jupiters using the 2m Himalayan Chandra 
Telescope (HCT), IAO and 1.3m Jagadish Chandra Bhattacharyya Telescope (JCBT), 
VBO to demonstrate the capability of these telescopes and their back-end 
instruments for precise transit photometry. The HCT has already been a part 
of the detection of the planet TRAPPIST-1 b \citep{gillon16}. This had 
motivated us to continue the search and follow-up observations using HCT 
and JCBT. In this paper we present the transit light curves of the hot
gas giants WASP-33 b, WASP-50 b, WASP-12 b, HATS-18 b and HAT-P-36 b.

Apart from the photometric noise, noise due to other sources such as stellar
activity, variation in the sky transparency etc. contribute to the
fluctuations in the transit light curves to a great extent which subsequently
constrain the precision in determining the transit parameters. Different types of noise require different treatments.
We have performed differential photometry to reduce the patterns that affect all the stars in a frame equally such as gradual variation in the airmass over the span of
3-4 hours of observation, large-scale transparency fluctuations that change the apparent brightness of all the stars in the frame almost equally etc. Also, prior to the modeling of the transit light curves, we have performed preprocessing on the light curves using the wavelet-based denoising technique. This decorrelation method suppresses the patterns of variability that are common to all the stars in the field but uncorrelated in time. These patterns can be caused by the medium-scale transparency fluctuations or seeing fluctuations that affect the stars on a frame slightly unequally but temporally uncorrelated. Also, this method removes the outliers caused by the cosmic-ray hits etc. Wavelet-based light curve noise analysis and filtering can be found in \cite{cubillos17,waldman14}. In case of transit photometry, \cite{ser18} has shown that this denoising technique does not alter the underlying astrophysical signals but improves the results obtained after modeling the transit light curves. Such preprocessing of data is a part of our self-developed pipeline used for reduction, analysis and modeling. For a comparison, we have also analysed the data without using the wavelet denoising process.

In addition to these patterns, confusing signals caused by stellar activity or pulsation are unique to each star and temporally correlated and can not be suppressed or removed by decorrelation or de-noising. They can, however, be modelled alongside the signal of interest. Such activity or pulsations have already been reported for some of the host stars \citep{vonessen14,mancini15}. We have addressed to such situations by modeling the covariance structure of the confusing signal using GP regression \citep{johnson15,barclay15}, in order to ensure that their effects are accounted for properly in the posterior uncertainty estimates for the fitted parameters of the transit light curves. Photon noise, propagated through the differential photometry, is uncorrelated which contributes to the diagonal elements of the covariance matrix. As a result of these processing and modeling we could determine the transit parameters of the planets more accurately than the previously published results.
We have adopted the stellar parameters such as mass, radius and effective temperature of the host stars and the semi-amplitude of the oscillation of the radial velocity of the host stars due to the planets ($K_{RV}$) from existing data available in the literature \citep{cameron10,gillon11, bakos12,lehmann15,penev16,collins17}.

The paper is organized in the following way: In section-\ref{sec:dodal} 
we discuss the details of the observations and the additional data 
adopted from the literature. In section-\ref{sec:rodam} we outline our
newly developed pipeline used for data reduction and analysis on the raw
images that produce the transit light curves that are fitted with transit
models. In section-\ref{sec:ton} we elaborate the preprocessing method
employed that reduces the fluctuations due to the contributions by various sources
of noise to the transit light curves.
In Section-\ref{sec:obc} we present the results for the observations of the
stars without planetary transit and discuss their utility 
in characterizing the stability of the baselines of the transit light curves.
In Section-\ref{sec:rd} we discuss the results and the work is summarized 
and concluded in section-\ref{sec:c}

\section{Details of Observations and Data Adopted from Literature} \label{sec:dodal}

We observed the transit events by using the 2-meter Himalayan Chandra 
Telescope (HCT) at Indian Astronomical Observatory (IAO), Hanle and the
1.3-meter Jagadish Chandra Bhattacharyya Telescope (JCBT) at Vainu Bappu
Observatory (VBO), Kavalur. For HCT we used the back-end instrument Hanle
Faint Object Spectroscopic Camera (HFOSC) which has a ${\rm 2K\times2k}$ 
optical CCD as the imager with a field of view of $10^\prime\times10^\prime$
on-sky. In case of JCBT we used the ${\rm 2k\times4k}$ UKATC optical CCD 
as the imager with a field of view of $10^\prime\times20^\prime$ on-sky.
Bessel V, R and I filters were used for the observations.
Both the imagers have a plate-scale of $3\arcsec$/pixel and both are 
liquid Nitrogen cooled to make the dark noise negligible. In order to obtain multiple transit light curves, each target has been observed repeatedly.
Some of the observed frames had to be discarded as they were affected by
either passing cloud or due to the condensation of water on the CCD. 

We have used the orbital period and the semi-amplitude of the radial velocity
of the host star from  the literature as mentioned in the next subsection (see Table~\ref{tab:lit}).
The rest of the planetary parameters are deduced by the detail analysis
and modeling of the observational data.

\subsection{WASP-33b}

WASP 33 b is a hot jupiter that orbits around the host star HD 15082. We observed this object for 5 transit events - 
one from HCT in V filter on 09 Dec 2017, two from JCBT in I filter on 05
Jan 2018, 27 Jan 2018 and the other two from JCBT in V filter on 26 Dec 
2018 and 06 Jan 2019. The host star is an A5 type star \citep{grenier99}.
It has a mass of $1.495\pm0.031$ $M_\odot$ and a radius of 1.444$\pm$0.034 
$R_\odot$ \citep{cameron10}. It is a $\delta$ Sct variable star
with a V mag of 8.3 \citep{herrero11}. So, the transit light curves of 
WASP-33b are contaminated with the pulsations as reported by 
\cite{vonessen14,johnson15}. The effect of these pulsations on the 
estimation of the transit parameters is subtracted by adopting a denoising
technique as explained in Sec-\ref{sec:ton}. The orbital period of 
the planet is taken as $1.21987\pm0.000001$ days \citep{cameron10,vonessen14,johnson15}.

In order to determine the mass and hence the mean density of the planet, we have
considered the semi-amplitude of the radial velocity of the star due 
to the planet as $K_{RV}$ = $304.0 \pm 20.0 {\rm ms^{-1}}$ \citep{lehmann15}.
The effective temperature of the host star is taken to be $T_{eff} = 7308 \pm 71 $ K \citep{cameron10}. From this the equilibrium temperature of the planet is determined (see section \ref{sec:rd}).

\subsection{WASP-50b}

We observed a total of 5 transit events of this hot jupiter by using JCBT 
in I filter on 26 Jan 2018, 28 Jan 2018 and 30 Jan 2018 and in R filter on
07 Jan 2019 and 11 Jan 2019. The host star has a V mag of 11.44, a mass of $0.892^{+0.080}_{-0.074}$ $M_\odot$ and a radius of $0.843\pm 0.031$ $R_\odot$ \citep{gillon11}. The $T_{eff}$ and the semi-amplitude
of the radial velocity of the star due to the planet ($K_{RV}$) are
respectively 5400$\pm$100 K and 256.6$\pm$4.4 m/s \citep{gillon11}. The
orbital period of the planet is taken as $1.955100\pm0.000005$ days \citep{gillon11}.

\subsection{WASP-12b}

 By using JCBT, we observed a total of 5 transit events  for this hot Jupiter -
3 in R-band on 03 Feb 2018, 04 Feb 2018 and 14 Feb 2018, one in I-band on 15
Feb 2018 and the other one in V-band on 04 Jan 2019. The host star has a 
mass, radius and $T_{eff}$ of 1.434$\pm$0.11 $M_\odot$, 1.657 $\pm 0.046$ 
$ R_\odot$ and 6300$\pm$150 K respectively \citep{collins17}. The semi-amplitude of the radial
velocity of the star due to the planet is $K_{RV}$ = 226$\pm4.0 {\rm 
ms^{-1}}$ \citep{collins17}. The orbital period of the planet is
$1.09142 \pm 0.00000014$ days \citep{collins17}.

\subsection{HATS-18b}

HATS-18 b is a hot Jupiter that orbits around a G type star which is very
similar to the Sun in terms of mass, radius and $T_{eff}$. We report the
observations of four transit events of HATS-18 b, all by using JCBT. The 
observations were taken in I-band on 27 Jan 2018, 18 Feb 2018 and 06 Apr 2018
and in R-band on 08 Jan 2019. The host star has a mass, radius and
 $T_{eff}$ of 1.037$\pm$0.047 $M_\odot$, $1.020^{+0.057}_{-0.031} R_\odot$ 
and 5600$\pm$120 K respectively \citep{penev16}. The semi-amplitude of radial velocity of
the star is $K_{RV}=415.2\pm10.0 {\rm ms^{-1}}$ \citep{penev16}. 
The orbital period of the planet is $0.8378\pm0.00000047$ days \citep{penev16}.

\subsection{HAT-P-36b}

We observed HAT-P-36b during 4 transit events: On 15 Feb 2018 by using I filter,
on 08 Apr 2018 and 06 May 2018 by using V filter in JCBT and on 20 Jun 2018
by using V filter in HCT.
The host star is a G5V star with mass, radius and $T_{eff}$ of $1.03\pm0.03$ 
$M_\odot$, $1.041\pm0.013$ $R_\odot$ and $5620\pm40$ K respectively \citep{bakos12}. 
This star is also very similar to the Sun in terms of mass, radius and 
$T_{eff}$. The semi-amplitude of radial velocity of the star due to the planet 
is $K_{RV}$ = $334.7\pm14.5 {\rm ms^{-1}}$ \citep{bakos12,mancini15}. 
The orbital period of the planet is taken as $1.9551\pm0.00005$ days \citep{bakos12,mancini15}.

\clearpage

\begin{deluxetable}{lccccc}[!ht]
\tablecaption{Stellar and orbital parameters adopted from literature \label{tab:lit}}
\tabletypesize{\scriptsize}
\tablehead{\\
Parameters            & WASP-33 b              & WASP-50 b                & WASP-12 b                        & HATS-18 b                     & HAT-P-36 b 
}
\startdata
Host star mass,       & $1.495\pm0.031$        & $0.892^{+0.08}_{-0.074}$ & $1.434\pm0.11$                   & $1.037\pm0.047$               & $1.03\pm0.03$             \\
$M_*$ ($M_\odot$)     &                        &                          &                                  &                               &                           \\        
Host star radius,     & $1.444\pm0.034$        & $0.843\pm0.031$          & $1.657\pm0.046$                  & $1.02^{+0.057}_{-0.031}$      & $1.041\pm0.013$           \\
$R_*$ ($R_\odot$)     &                        &                          &                                  &                               &                           \\
Host star $T_{eff}$   & $7430\pm100$    (a)    & $5400\pm100$             & $6360\pm140$                     & $5600\pm120$                  & $5620\pm40$               \\
 (K)                  &                        &                          &                                  &                               &                           \\
Orbital Period,       & $1.21987\pm0.000001$   & $1.955100\pm0.000005$    & $1.09142\pm1.4432\times 10^{-7}$ & $0.83784\pm4.7\times 10^{-7}$ & $1.32734683\pm0.00000048$ \\
P (days)              &                        &                          &                                  &                               &                           \\
RV amplitude,         & $304\pm20$             & $256.6\pm4.4$            & $226.4\pm4.1$                    & $415.2 10.0$                  & $334.7\pm14.5$            \\
$K_{RV}$ (m/s)        &                        &                          &                                  &                               &                           \\
Sources               & \cite{cameron10}       & \cite{gillon11}          & \cite{collins17}                 & \cite{penev16}                & \cite{bakos12}            \\
                      & \cite{lehmann15}       &                          &                                  &                               & \cite{mancini15}          \\                                                   
\enddata
\tablecomments{The value of each parameter is shown along with 1-$\sigma$ error margin.}
\end{deluxetable}

\clearpage

\clearpage

\begin{deluxetable}{lcccccccc}[!ht]
\tablecaption{Observation details and the night-dependent model parameters \label{tab:obs}}
\tabletypesize{\scriptsize}
\tablehead{\\
	Planet    & Date of     & Telescope & Filter   & Photometric  & Mid-transit ephemerides,     & Cycle                  & $A$                              & $\tau$       \\ 
              & Observation &           & (Bessel) & S/N (median) & $t_{cen}$ ($BJD-TDB$)        & no.\tablenotemark{a}   &                                  &     
}
\startdata 
              & 09 Dec 2017 & HCT       & V        & 191.90       & $2458097.30431\pm0.00000786$ & 4191                   & $0.0060\pm0.0001$                & $20.0\pm0.1$  \\
              & 05 Jan 2018 & JCBT      & I        & 1253.47      & $2458124.14196\pm0.00000769$ & 4213                   & $0.0017\pm0.0001$                & $20.0\pm0.1$  \\
WASP-33 b     & 27 Jan 2018 & JCBT      & I        & 300.29       & $2458146.09962\pm0.00000765$ & 4231                   & $0.0029\pm0.0001$                & $19.99\pm0.1$ \\ 
              & 26 Dec 2019 & JCBT      & V        & 505.03       & $2458479.12739\pm0.00000698$ & 4504                   & $0.0044\pm0.0001$                & $19.99\pm0.1$ \\
			  & 06 Jan 2019 & JCBT      & V        & 466.48       & $2458490.10444\pm0.00000696$ & 4513                   & $0.0039\pm0.0001$                & $18.99\pm0.1$ \\
\hline                                                                                                                    
              & 26 Jan 2018 & JCBT      & I        & 361.93       & $2458145.20327\pm0.00001057$ & 1323                   & $0.00141\pm0.0001$               & $12.0\pm0.1$  \\
              & 28 Jan 2018 & JCBT      & I        & 252.39       & $2458147.15848\pm0.00001135$ & 1324                   & $0.00232\pm0.0001$               & $13.0\pm0.1$ \\
WASP-50 b     & 30 Jan 2018 & JCBT      & I        & 1012.89      & $2458149.11405\pm0.00000886$ & 1325                   & $0.00096\pm0.0001$               & $10.0\pm0.1$  \\ 
              & 07 Jan 2019 & JCBT      & R        & 1139.06      & $2458491.25582\pm0.00000783$ & 1500                   & $0.00240\pm0.00011$              & $12.0\pm0.1$   \\
              & 11 Jan 2019 & JCBT      & R        & 1086.67      & $2458495.16601\pm0.00000811$ & 1502                   & $0.00240\pm0.00011$              & $12.0\pm0.1$   \\
\hline                                                                                                                     
              & 03 Feb 2017 & JCBT      & R        & 1318.76      & $2458153.22835\pm0.00001885$ & 2754                   & $0.00037^{+0.0005}_{-0.0001}$    & $12.0\pm0.1$   \\
              & 04 Feb 2018 & JCBT      & R        & 1180.83      & $2458154.31975\pm0.00000770$ & 2755                   & $0.00302\pm0.0002$               & $10.0\pm0.1$  \\
WASP-12 b     & 14 Feb 2018 & JCBT      & R        & 1261.32      & $2458164.14255\pm0.00002867$ & 2764                   & $0.00100^{+0.00025}_{-0.00014}$  & $10.0\pm0.1$  \\
              & 15 Feb 2018 & JCBT      & I        & 500.50       & $2458165.23478\pm0.00001119$ & 2765                   & $0.00293^{+0.0001}_{-0.0004}$    & $13.0\pm0.1$  \\ 
              & 04 Jan 2019 & JCBT      & V        & 1085.84      & $2458488.29489\pm0.00001192$ & 3646                   & $0.00300^{+0.0007}_{-0.0001}$    & $12.0\pm0.1$   \\
\hline                                                                                                                    
              & 27 Jan 2018 & JCBT      & I        & 443.07       & $2458146.42651\pm0.00000928$ & 1261                   & $0.00099\pm0.0001$               & $12.0\pm0.1$  \\
    HATS-18 b & 18 Feb 2018 & JCBT      & I        & 239.80       & $2458168.21044\pm0.00000952$ & 1287                   & $0.004\pm0.0001$ 			     & $10.0\pm0.1$  \\
              & 06 Apr 2018 & JCBT      & I        & 307.09       & $2458215.12967\pm0.00001042$ & 1343                   & $0.00599\pm0.0001$	             & $7.0\pm0.1$   \\ 
              & 08 Jan 2019 & JCBT      & R        & 425.84       & $2458492.45583\pm0.00001028$ & 1674                   & $0.00091\pm0.0001$               & $7.0\pm0.1$   \\
\hline                                                                                                                     
              & 15 Feb 2017 & JCBT      & I        & 329.09       & $2458165.45507\pm0.00000686$ & 1959                   & $0.00301\pm0.0001$               & $13.0\pm0.1$  \\
   HAT-P-36 b & 08 Apr 2018 & JCBT      & V        & 461.99       & $2458217.22160\pm0.00000755$ & 1998                   & $0.00175\pm0.0001$               & $12.99\pm0.1$ \\
              & 06 May 2018 & JCBT      & V        & 589.10       & $2458245.09464\pm0.00000737$ & 2019                   & $0.00099\pm0.0001$ 			     & $13.0\pm0.1$  \\
              & 20 Jun 2018 & HCT       & V        & 1106.84      & $2458290.22569\pm0.00000709$ & 2053                   & $0.00088\pm0.0001$    		     & $13.0\pm0.1$  \\
\enddata 
\tablenotetext{\text{a}}{The mid-transit ephemerides (BJD-TDB) at cycle 0 for WASP-33 b, WASP-50 b, WASP-12 b, HATS-18 b, HAT-P-36 b are considered to be at 2452984.82964 \citep{turner16}, 2455558.61197 \citep{gillon11}, 2455147.4582 \citep{turner16}, 2457089.90598 \citep{penev16} and 2455565.18167 \citep{mancini15} respectively.}
\tablecomments{The values of $t_{cen}$, A and $\tau$ are shown along with 1-$\sigma$ error margin.}
\end{deluxetable}

\clearpage

\section{Data Reduction, Analysis and Modeling} \label{sec:rodam}

We have developed an automated pipeline based on Python and PyRAF to
reduce, analyse and model the observed data. 
This pipeline performs the necessary bias and flat corrections on the raw 
images and then aligns those corrected images. Leveraging the moderately
large field of view on-sky of the imagers we could image the target stars
along with a few field stars that serves as reference stars for those 
targets. Subsequently, using the DAOPHOT package of PyRAF
 the pipeline automatically performs differential aperture photometry on the
target stars and generates normalized light curves during the transit ephoch.
Henceforth, by a single light curve we will imply the light curve of a host 
star obtained on one particular night which may or may not contain a transit
signal.

 The light curves were denoised using wavelet denoising technique before modeling to decorrelate the patterns of variability common to all the stars in the frame but uncorrelated in time \citep{ser18}. Also, to take care of the noise pattern unique to the host stars and correlated in time caused by stellar activity or pulsation etc.,we have adopted Gaussian process correlated noise modeling technique during modeling to model its covariance structure and to take it into account when calculating the likelihood of the data given the model \citep{johnson15,barclay15}. The various sources of noise that causes the fluctuations in the light curves and their decorrelation or modeling techniques are discussed in details in the next section.

After denoising, we modeled the transit light curves by using the formalisms
described in \cite{mandel-agol02}. We have used the Markov Chain Monte Carlo
(MCMC) technique employing Metropolis-Hastings algorithm \citep{cameron10} to fit the models with the observed light curves and thus determined the various physical parameters from the best fit. An essential
parameter of the model is the orbital period which we have kept fixed at 
the values given in previously published results \citep{cameron10,gillon11,
bakos12,penev16,collins17}. For all the transit events we have assumed 
circular orbits of the planets. The free parameters for each transit model
are the mid-transit ephemeris ($t_{cen}$), impact parameter $(b)$, the
scaled radius of the star ($R_*/a$), the ratio between the planetary and the
stellar radius ($R_p/R_*$), the pre-ingress or post-egress baseline level
($f_{star}$) and the limb darkening coefficients ($Ci$). We have modeled
all the observed transit light curves of a particular planet simultaneously.
By modeling the light curves simultaneously we have deduced a single set of values for $b$, $R_*/a$ and
$R_p/R_*$ for each planet. These parameters are the properties of the planet-star systems and hence independent of the observing conditions.

We deduced different sets of values for the limb darkening coefficients $C_i$
for each host star for different filters. Also, for different transit events, 
we deduced different sets of values for $t_{cen}$ and $f_{star}$ from our model
fit (Table-\ref{tab:obs}) as these parameters depend on the nights of observations. For all the free parameters other than the limb darkening coefficients we have set uniform prior function \citep{gillon11}.
We adopted quadratic limb darkening law which can be expressed as:
\begin{equation}
I/I(\mu=1) = 1 - C_1(1-\mu) - C_2(1-\mu^2),
\end{equation}
where $I/I(\mu=1)$ denotes the intensity at any point on the disc normalized 
to that at the center. The initial values required to derive the limb darkening coefficients from the MCMC fit are taken from \cite{claret11} and Gaussian priors was set on them \citep{johnson15}.

The MCMC generates a sample space of the best-fit values for the model
parameters depending upon the number of walkers and iterations by maximizing 
the likelihood space of model fits to the light curve data. A Gaussian fit
to the sample space then gives the required value of the parameters at 
 $1 \sigma$ error margins.

\section{Treatment of Noise} \label{sec:ton}

The images captured from  ground based telescopes are susceptible to noises generated from various sources. These noises are either common to all the objects in a frame and uncorrelated in time such as the noises caused by the fluctuations in the transparency, seeing, airmass etc. or unique to each object and correlated in time such as the the noises caused by the activity or pulsations of the host stars. In order to decorrelate the former type of noises from the light curves, a preprocessing on the light curves is essential before modeling to achieve high precision in the transit parameters estimated from modeling. However, the smoothing techniques such as Moving Average or Gaussian smoothing can not be used to suppress these noises as the smoothing process can distort the orginal light curves by removing the high frequency components of the transit 
signal itself and question the reliability of the properties derived therefrom.

On the other hand, for a non-stationary non-sinusoidal signal like a noisy transit signal, the wavelet denoising is much more efficient than a frequency-based filtering technique in terms of signal reconstruction and denoised S/N \citep{barsanti11,lagha13}. Wavelets have already been used extensively in the light curve noise analysis and filtering \citep{cubillos17,waldman14}. In case of transit photometry, wavelet denoising can efficiently remove the outliers, yield better MCMC posterior distributions and reduce the bias in the fitted transit parameters and their uncertainties \citep{ser18}. We used the pywt package \citep{pywt} and followed the same procedure as described in \cite{ser18}. Also, we simulated a transit light curve assuming a set of values for the transit parameters along with uncertainties in each parameter. The uncertainties in the parameters then reflect the errorbars in the simulated transit light curve. The wavelet denoising process is expected not to affect a light curve with errorbars limited by the uncertainty in the transit parameters. In fact we found that our simulated transit light curve was almost unchanged by the denoising process. This ensures that the light curves are not over-smoothed or the errorbars are not under-estimated by the denoising process.

\begin{figure}[!ht]
\centering
\includegraphics[scale=0.45,angle=0]{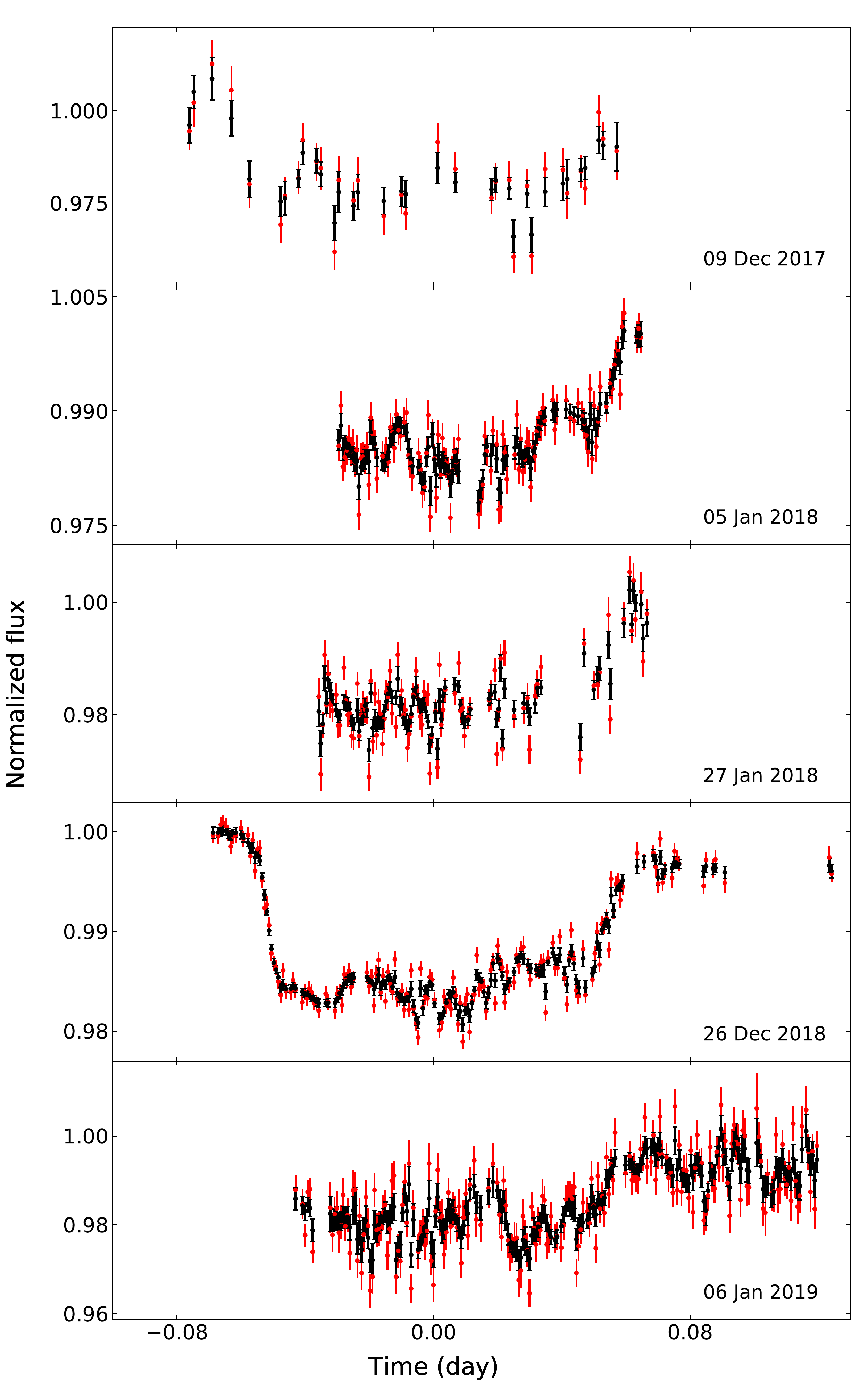}
\caption{The transit light curves of WASP-33b. The original light curves (right after differential photometry) are shown with red errorbars and wavelet denoised light curves are over-plotted with black errorbars. The zero points on the time axes are set at the the mid-transit ephemerides as shown in Table~\ref{tab:obs}.
\label{fig:wasp33bwd}}
\end{figure}

\begin{figure}[!ht]
\centering
\includegraphics[scale=0.45,angle=0]{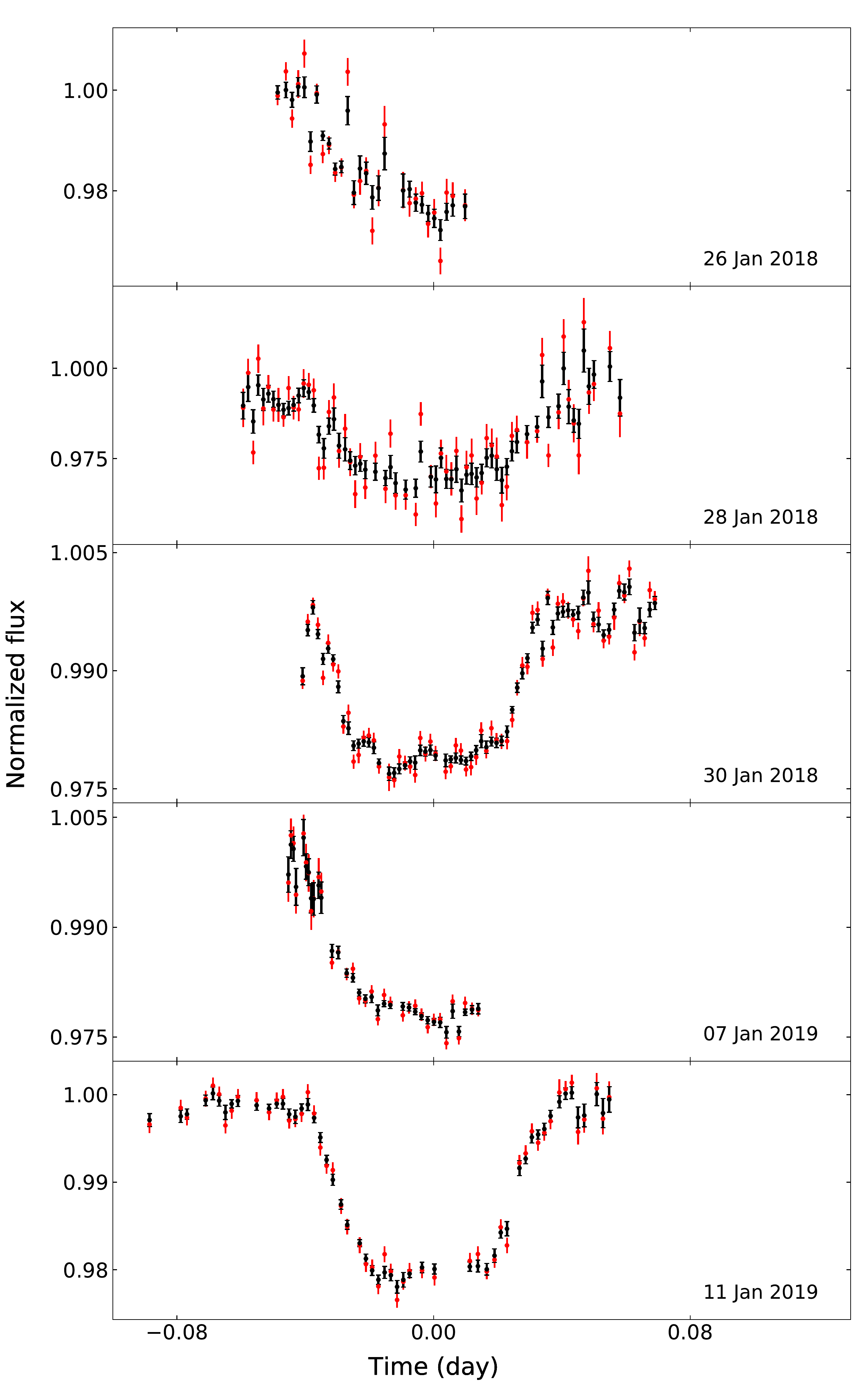}
\caption{The transit light curves of WASP-50b. The original light curves (right after differential photometry) are shown with red errorbars and wavelet denoised light curves are over-plotted with black errorbars. The zero points on the time axes are set at the the mid-transit ephemerides as shown in Table~\ref{tab:obs}.
\label{fig:wasp50bwd}}
\end{figure}

\begin{figure}[!ht]
\centering
\includegraphics[scale=0.45,angle=0]{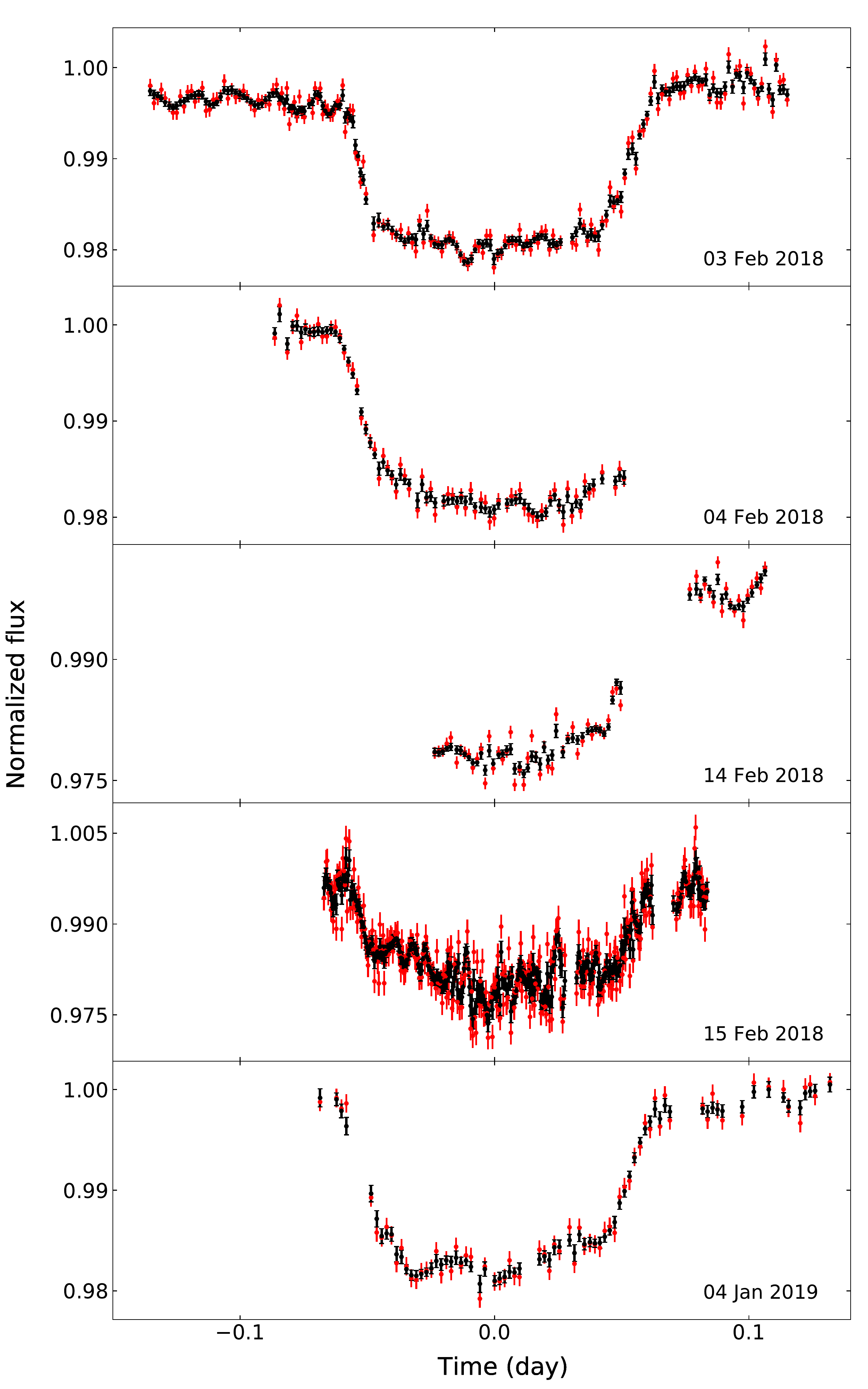}
\caption{The transit light curves of WASP-12b. The original light curves (right after differential photometry) are shown with red errorbars and wavelet denoised light curves are over-plotted with black errorbars. The zero points on the time axes are set at the the mid-transit ephemerides as shown in Table~\ref{tab:obs}.
\label{fig:wasp12bwd}}
\end{figure}

\begin{figure}[!ht]
\centering
\includegraphics[scale=0.45,angle=0]{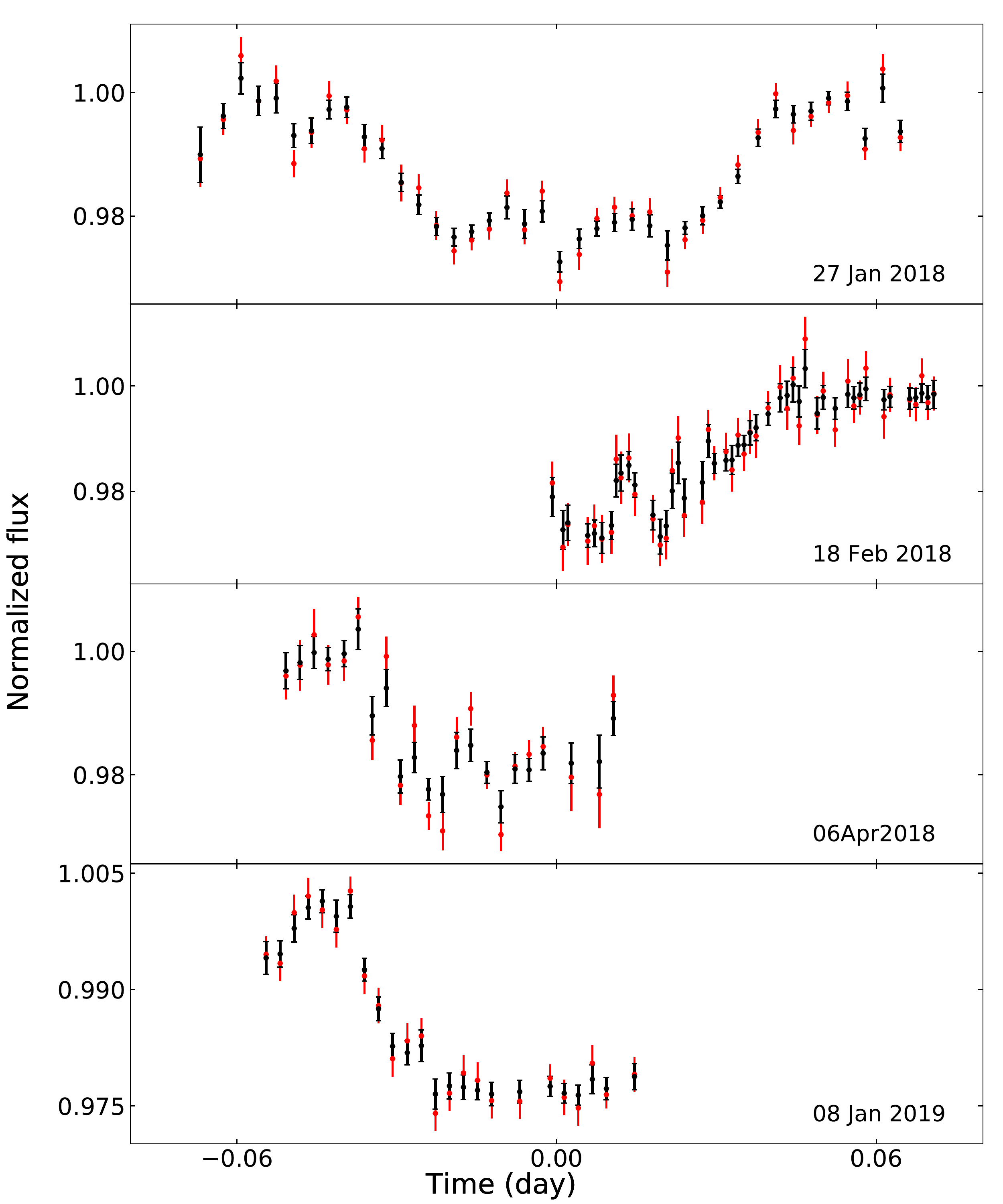}
\caption{The transit light curves of HATS-18b. The original light curves (right after differential photometry) are shown with red errorbars and wavelet denoised light curves are over-plotted with black errorbars. The zero points on the time axes are set at the the mid-transit ephemerides as shown in Table~\ref{tab:obs}.
\label{fig:hats18bwd}}
\end{figure}

\begin{figure}[!ht]
\centering
\includegraphics[scale=0.45,angle=0]{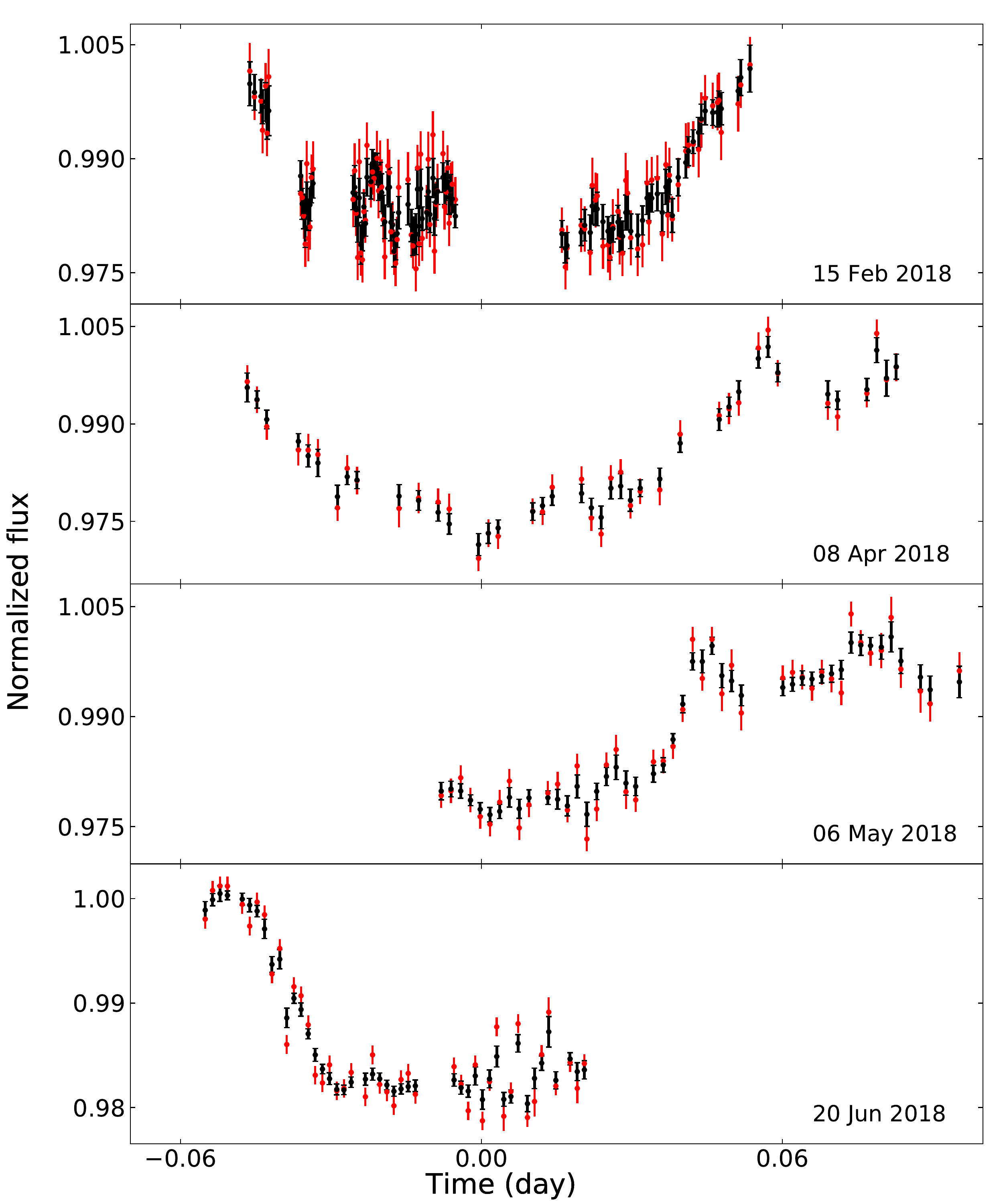}
\caption{The transit light curves of HAT-P-36b. The original light curves (right after differential photometry) are shown with red errorbars and wavelet denoised light curves are over-plotted with black errorbars. The zero points on the time axes are set at the the mid-transit ephemerides as shown in Table~\ref{tab:obs}.
\label{fig:hatp36bwd}}
\end{figure}

The transit light curves with and without wavelet-denoising are shown in 
Figure~\ref{fig:wasp33bwd}, Figure~\ref{fig:wasp50bwd}, Figure~\ref{fig:wasp12bwd}, Figure~\ref{fig:hats18bwd} 
and in Figure~\ref{fig:hatp36bwd}. The values of the planetary physical parameters deduced by modeling the transit light curves preprocessed with wavelet denoising are presented in Table~\ref{tab:par}. The same without wavelet denoising process are provided in Table~\ref{tab:parnw}. A comparison of the results presented in the two tables implies that the wavelet denoising process improves the precision in the deduced parameters significantly. However, the wavelet denoising process can only efficiently remove the outliers and reduce the temporally uncorrelated noise. The temporally correlated noise unique to the host stars still remaining in the denoised transit light curves can not be directly decorrelated from the light curves. Instead, while modeling the light curves using MCMC, we also modeled the covariance structure of the correlated noise using the Gaussian process regression (GP) \citep{johnson15,barclay15} whose mean function includes the transit model function \citep{mandel-agol02} itself. This ensures that the effect of the correlated noise on the estimation of the posterior uncertainties on the fitted parameters is minimized. The diagonal elements of the covariance matrix are contributed by the errorbars in the transit light curves (photon noise plus readout noise). We have formed the correlation matrix using a Matern 3/2 kernel, given by \citep{johnson15}:
\begin{equation}
\kappa_{ij} = A^2(1+\frac{\sqrt{3}\Delta t_{ij}}{\tau})\exp(-\frac{\sqrt{3}\Delta t_{ij}}{\tau}) + \delta_{ij}\sigma^2_i,
\end{equation}
where, $\Delta t_{ij} = (t_i-t_j)$; $t_i,t_j$ are two points of time of
observation, $\sigma_i$ is the uncertainty (error) in the flux values at
time $t_i$ and $\delta_{ij}$ is the Kronecker delta function. 
$A$ and $\tau$ are the amplitude and the time scale of the fluctuation of a
light curve due to the correlated noise and used in the MCMC for model fitting.
We have kept $A$ and $\tau$ variable for each light curve. The prior
functions of $A$ and $\tau$ are also chosen to be uniform. The prior function
for $A $ is estimated from the amplitude of fluctuation at the pre-ingress 
or post-egress points of time and the prior function of $\tau$ is
estimated from the high frequency peaks on the Lomb-Scargle periodogram 
of each light curve. Besides, while modeling, we have multiplied the
transit plus noise model with a baseline function to represent the 
systematics due to the other astrophysical noise or noise at the detector 
stage. By minimizing the Bayesian Information Criterion (BIC) we have 
chosen a one order baseline function of time for this purpose \citep{gillon16}. 

%\vspace{0.1\textwidth}
\clearpage 
\begin{deluxetable}{lccccc}[!ht] 
\tablecaption{Physical parameters directly obtained and deduced from our differential transit photometry followed by preprocessing with WD technique and modeling. \label{tab:par}}
\tablehead{\\
Parameters                                          & WASP-33 b             & WASP-50 b                  & WASP-12 b                       & HATS-18 b              & HAT-P-36 b
}
\startdata
\textbf{Transit model parameters} \\                                                                                             
Impact Parameter, b                                 & $0.21\pm0.002$        & $0.669^{+0.018}_{-0.007}$  & $0.339\pm0.0017$                & $0.3\pm0.001$          & $0.25\pm0.007$               \\        
Scaled Stellar radius, $R_*/a$                      & $0.28\pm0.0008$       & $0.133\pm0.003$            & $0.333^{+0.0002}_{-0.0017}$     & $0.273\pm0.0006$       & $0.21^{+0.003}_{-0.0002}$    \\
Planet/Star Radius Ratio, $R_p/R_*$                 & $0.1118\pm0.0002$     & $0.139\pm0.0006$           & $0.117\pm0.0002$                & $0.132\pm0.0004$       & $0.1199\pm0.0002$            \\
\hline
\textbf{Limb darkening coefficients} \\
Linear Term for V filter, $C1_V$                    & $0.5\pm0.01$          & $-$                        & $0.42\pm0.01$                   & $0.5\pm0.01$           & $0.53\pm0.01$                \\
Quadratic Term for V filter, $C2_V$                 & $0.2\pm0.01$          & $-$                        & $0.31\pm0.01$                   & $0.2\pm0.01$           & $0.23\pm0.01$                \\
Linear Term for R filter, $C1_R$                    & $-$                   & $0.4\pm0.01$               & $0.3\pm0.01$                    & $0.41\pm0.01$          & $-$                         \\
Quadratic Term for R filter, $C2_R$                 & $-$                   & $0.2\pm0.01$               & $0.3\pm0.01$                    & $0.18\pm0.01$          & $-$                         \\
Linear Term for I filter, $C1_I$                    & $0.31\pm0.01$         & $0.3\pm0.01$               & $0.29\pm0.01$                   & $0.31\pm0.01$          & $0.32\pm0.01$                \\
Quadratic Term for I filter, $C2_I$                 & $0.18\pm0.01$         & $0.2\pm0.01$               & $0.31\pm0.01$                   & $0.21\pm0.01$          & $0.19\pm0.01$                \\
\hline
\textbf{Deduced parameters}\\
Transit Duration, $T_{14}$ (days)                   & $0.1189\pm0.0005$      & $0.0764\pm0.0011$          & $0.1267^{+0.00009}_{-0.0005}$   & $0.081\pm0.0001$       & $0.093^{+0.0016}_{-0.00007}$ \\
Planet Radius, $R_p$ ($R_J$)                        & $1.593\pm0.074$       & $1.166\pm0.043$            & $1.937\pm0.056$                 & $1.329\pm0.075$        & $1.277\pm0.02$               \\
Scale Parameter, $a/R_*$							& $3.571\pm0.01$	    & $7.51\pm0.10$	             & $3.0^{+0.016}_{-0.0019}$        & $3.658\pm0.008$        & $4.95\pm0.042$               \\ 
Orbital Separation, $a$ (AU)                     	& $0.0239\pm0.00063$    & $0.0293\pm0.0013$          & $0.0232\pm0.00064$              & $0.0174\pm0.00098$     & $0.0241\pm0.00047$           \\
Orbital Inclination, $i$ (degrees)                  & $86.63\pm0.03$        & $84.88\pm0.27$             & $83.52\pm0.03$                 & $85.29\pm0.013$        & $87.13^{+0.004}_{-0.13}$     \\
Planet Mass, $M_p$ ($M_J$)                          & $2.093\pm0.139$       & $1.4688\pm0.092$           & $1.465\pm0.079$                 & $1.9795\pm0.076$       & $1.8482\pm0.087$             \\
Planet Mean Density, $\rho_p$ (g$cm^{-3}$)          & $0.689\pm0.074$       & $1.325\pm0.214$            & $0.267\pm0.0288$                & $1.1169\pm0.216$         & $1.175\pm0.078$              \\
Surface Gravity, $\log g_p$ (cgs)                   & $3.275\pm0.04$        & $3.469\pm0.029$            & $2.998\pm0.01$                  & $3.45\pm0.013$         & $3.476\pm0.027$              \\
Equilibrium Temp.\tablenotemark{a}, $T_{eq}$ (K)    & $2781.70\pm41.1$      & $1394.84\pm32.7$           & $2592.6\pm57.2$                 & $2069.48\pm45.0$       & $1780.97\pm18.8$             \\
\enddata
\tablenotetext{\text{a}}{Assuming zero Bond albedo and full re-distribution of the incident stellar
flux.}
\tablecomments{The value of each parameter is shown along with 1-$\sigma$ error margin. Also, some of the limb darkening coefficients are shown as $-$, which implies that no transit has been observed for that particular planet in that filter.}
\end{deluxetable}

\clearpage

\begin{deluxetable}{lccccc}[!ht] 
\tablecaption{Physical parameters directly obtained and deduced from our differential transit photometry followed by modeling and no preprocessing (without WD). \label{tab:parnw}}
\tablehead{\\
Parameters                                          & WASP-33 b            & WASP-50 b                  & WASP-12 b                       & HATS-18 b                   & HAT-P-36 b
}
\startdata
\textbf{Transit model parameters} \\                                                                                             
Impact Parameter, b                                 & $0.21\pm0.003$      & $0.65^{+0.068}_{-0.005}$   & $0.339\pm0.007$               & $0.299\pm0.019$               & $0.247\pm0.02$       \\        
Scaled Stellar radius, $R_*/a$                      & $0.28\pm0.003$       & $0.133^{+0.01}_{-0.002}$   & $0.332\pm0.002$               & $0.26\pm0.005$                & $0.202\pm0.004$      \\
Planet/Star Radius Ratio, $R_p/R_*$                 & $0.1119\pm0.003$     & $0.135\pm0.001$            & $0.117^{+0.002}_{-0.0002}$    & $0.131^{+0.003}_{-0.0002}$    & $0.1199\pm0.003$     \\
\hline
\textbf{Limb darkening coefficients} \\
Linear Term for V filter, $C1_V$                    & $0.5\pm0.03$         & $-$                        & $0.4\pm0.04$                  & $0.48\pm0.04$                 & $0.5\pm0.05$         \\
Quadratic Term for V filter, $C2_V$                 & $0.2\pm0.03$         & $-$                        & $0.3\pm0.04$                  & $0.2\pm0.05$                  & $0.2\pm0.04$         \\
Linear Term for R filter, $C1_R$                    & $-$                  & $0.39\pm0.05$              & $0.3\pm0.05$                  & $0.4\pm0.06$                  & $-$                  \\
Quadratic Term for R filter, $C2_R$                 & $-$                  & $0.21\pm0.05$              & $0.3\pm0.05$                  & $0.21\pm0.04$                 & $-$                  \\
Linear Term for I filter, $C1_I$                    & $0.3\pm0.04$         & $0.3\pm0.06$               & $0.29\pm0.03$                 & $0.31\pm0.05$                 & $0.3\pm0.06$         \\
Quadratic Term for I filter, $C2_I$                 & $0.2\pm0.04$         & $0.2\pm0.06$               & $0.3\pm0.03$                  & $0.2\pm0.05$                  & $0.2\pm0.06$         \\
\hline
\textbf{Deduced parameters}\\
Transit Duration, $T_{14}$ (days)                   & $0.1188\pm0.0012$    & $0.078\pm0.003$            & $0.1267\pm0.0006$             & $0.079\pm0.0014$              & $0.095\pm0.0018$     \\
Planet Radius, $R_p$ ($R_J$)                        & $1.601\pm0.057$      & $1.144\pm0.057$            & $1.939\pm0.058$               & $1.341\pm0.079$               & $1.30\pm0.03$        \\
Scale Parameter, $a/R_*$							& $3.571\pm0.04$	   & $7.485^{+0.1}_{-0.63}$     & $3.0\pm0.019$                 & $3.724\pm0.067$               & $4.937\pm0.1$        \\ 
Orbital Separation, $a$ (AU)                     	& $0.0239\pm0.00071$   & $0.0289\pm0.002$           & $0.0231\pm0.00068$            & $0.0176\pm0.001$              & $0.0239\pm0.00058$   \\
Orbital Inclination, $i$ (degrees)                  & $86.6\pm0.05$        & $85.01^{+0.09}_{-1.01}$    & $83.52^{+0.08}_{-0.16}$       & $85.38\pm0.33$                & $87.13\pm0.23$       \\
Planet Mass, $M_p$ ($M_J$)                          & $2.093\pm0.1404$     & $1.4692\pm0.092$           & $1.465\pm0.079$               & $1.9794\pm0.077$              & $1.848\pm0.088$      \\
Planet Mean Density, $\rho_p$ (g$cm^{-3}$)          & $0.6774\pm0.095$     & $1.2958\pm0.21$            & $0.266\pm0.029$               & $1.088\pm0.217$               & $1.042\pm0.09$       \\
Surface Gravity, $\log g_p$ (cgs)                   & $3.268\pm0.037$      & $3.463^{+0.02}_{-0.1}$     & $2.99\pm0.015$                & $3.46\pm0.03$                 & $3.432\pm0.036$      \\
Equilibrium Temp.\tablenotemark{a}, $T_{eq}$ (K)    & $2784.09\pm45.9$     & $1404.63\pm58.3$           & $2596.19\pm58.1$              & $2052.09\pm51.4$              & $1789.92\pm23.2$     \\
\enddata
\tablenotetext{\text{a}}{Assuming zero Bond albedo and full re-distribution of the incident stellar
flux.}
\tablecomments{The value of each parameter is shown along with 1-$\sigma$ error margin. Also, some of the limb darkening coefficients are shown as $-$, which implies that no transit has been observed for that particular planet in that filter.}
\end{deluxetable}

\clearpage

\clearpage

\section{Observations for Baseline Characterization} \label{sec:obc}

A key element of the capability testing of the telescopes used for transit
observations is the proper characterization of the baseline of the light
curves that ensures the precision in the transit parameters. For this purpose,
we have observed some of the host stars out of transit. Also, we monitored  
some fields containing multiple stars for several days which are not known to harbour any 
confirmed or candidate planet yet. 

\subsection{Out-of-transit observations of the host stars}
\begin{figure}
\centering 
% \leavevmode 
% \setlength{\plot@width}{0.425\linewidth}
 \includegraphics[scale=0.38,angle=0]{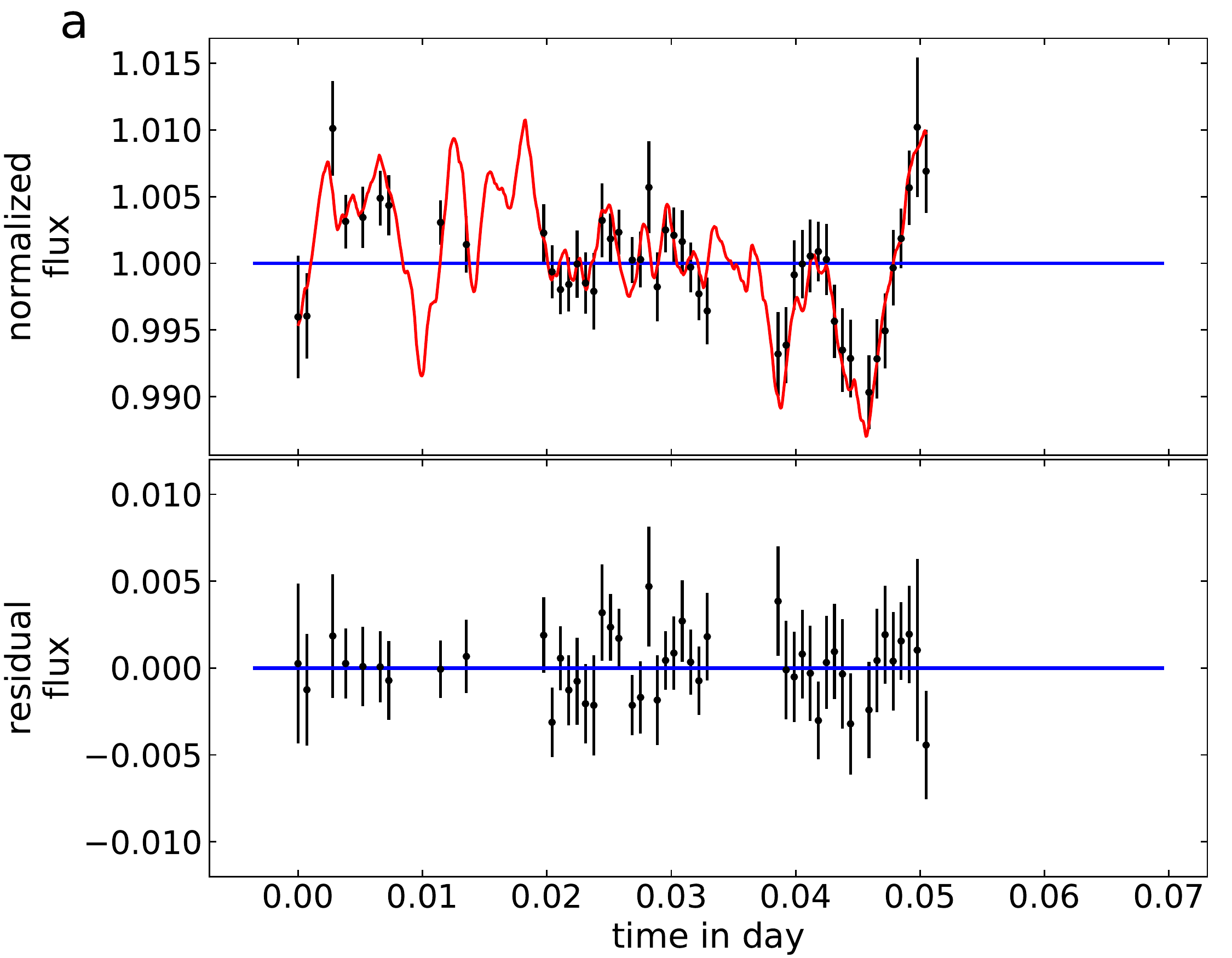}% 
 \hfil 
 \includegraphics[scale=0.38,angle=0]{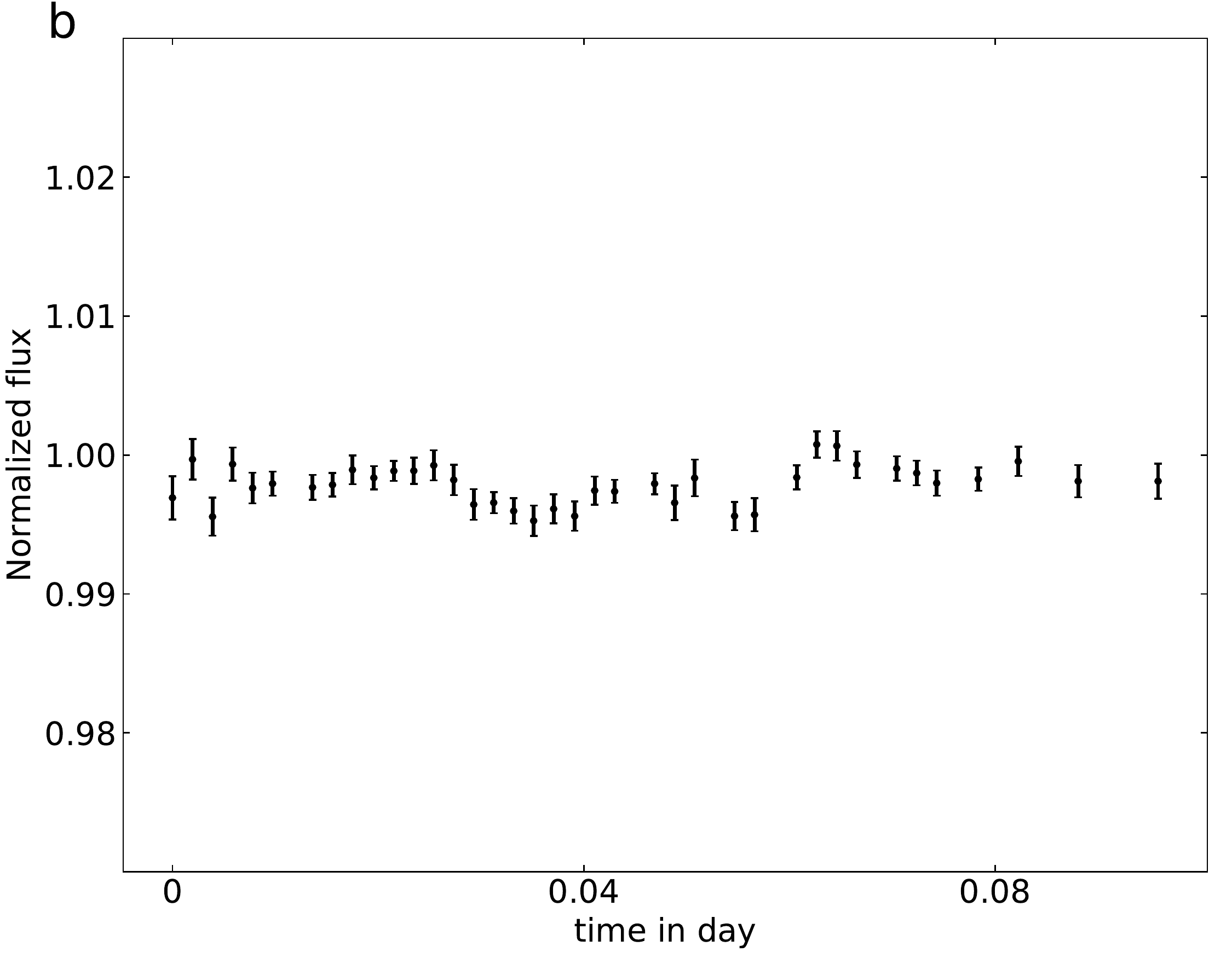}% 
%\gridline{\fig{wasp33nt.pdf}{0.47\textwidth}{}
%		  \fig{wasp50nt.pdf}{0.47\textwidth}{}
%		  }
\caption{a - Light curve of WASP-33 observed on 10 Jan 2019 from JCBT, when there was no predicted transit event. The zero point on the time axis is set at 2458494.074141 BJD-TDB. Top-The black error-bars represent the flux and error values obtained right after wavelet denoising. The red line denotes the GP noise model. Bottom- The black error-bars represent the residual flux after subtracting the GP noise model from the wavelet denoised flux values. b - Pre-ingress wavelet denoised light curve of WASP-50 observed on 09 Jan 2019 from JCBT. The zero point on the time axis is set at 2458493.060419156 BJD-TDB. None of the plots on either side show any detectable transit signature as expected.
\label{fig:waspnt}}
\end{figure}

In this paper we also report the results of the out-of-transit photometric observations of the
host stars WASP-33 and WASP-50. We observed 
WASP-33 from JCBT on 10 Jan 2019 in V-band  at 13:30-15 UT when there was no 
planetary transit. Similarly, we observed WASP-50 from JCBT on 09 Jan 2019
in R-band at 13:20-15:40 UT right before the transit by its planet.
After wavelet denoising, we found a fluctuation of $0.5\%$ in the light
curve of WASP-33 which after the subtraction of the GP regression model was reduced to 
$0.3\%$. The peak signal-to-noise ratio (PSNR) was improved from about 
200 to about 300 after the modeling. The light curves of 
WASP-50 shows a fluctuation of $0.1\%$ which required no noise modeling 
because modeling does not improve it further (Table~\ref{tab:base}).
The out-of-transit light curves of WASP-33 and WASP-50 during out-transit epoch are shown in 
Figure~\ref{fig:waspnt}.

\subsection{Observations of stars with no known planet}
\begin{figure}
\centering
\includegraphics[scale=0.38,angle=0]{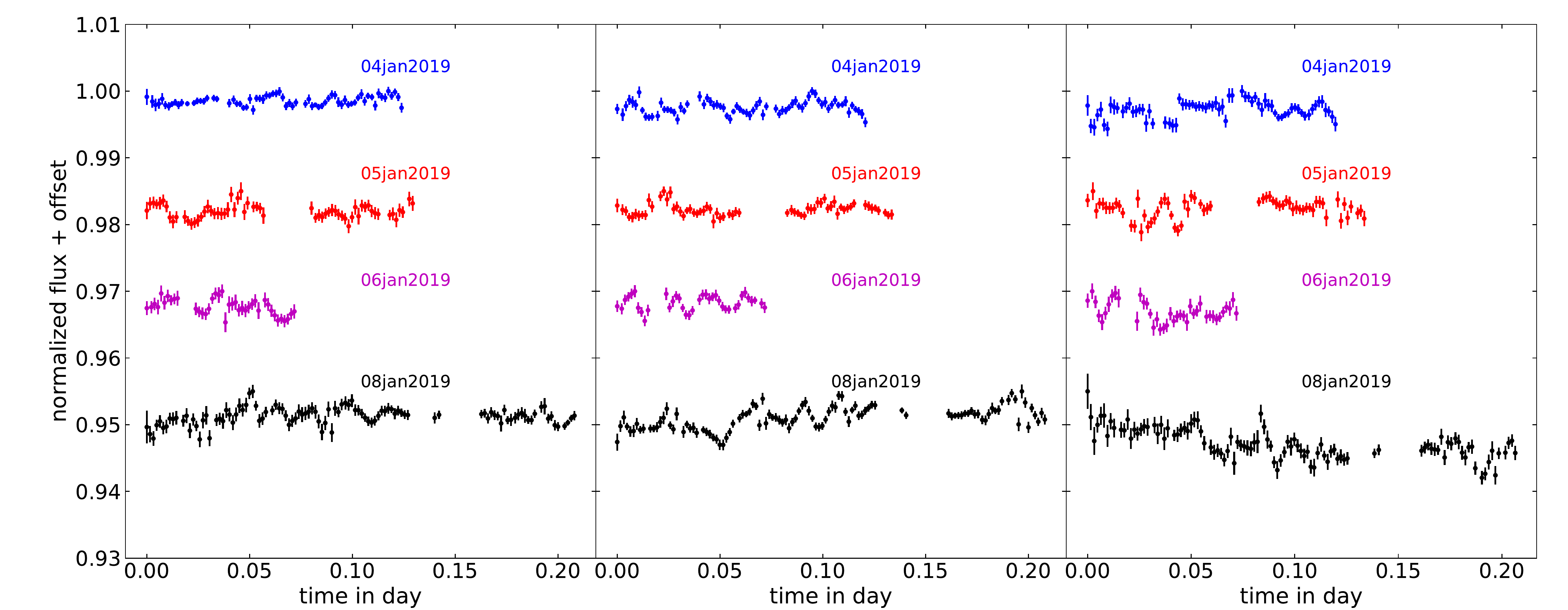}
%\gridline{\fig{wasp33nt.pdf}{0.47\textwidth}{}
%		  \fig{wasp50nt.pdf}{0.47\textwidth}{}
%		  }
\caption{The light curves of the stars TYC 3337-1778-1, TYC 3337-1676-1 and TYC 3337-83-1 (from left to right) respectively observed from JCBT. The zero points on the time axes for the dates 04 Mar 2019, 05 Jan 2019, 06 Jan 2019 and 08 Mar 2019 are set at BJD-TDB  2458488.096490033, 2458489.094740324, 2458490.231318539 and 2458492.054523 respectively.
\label{fig:ngc1545}}
\end{figure}

We observed the field around the star TYC 3337-1778-1. No star in 
this field is reported yet to have any planet. This field was neither
a part of the Kepler or K2 survey nor has it been included in the list 
of candidates for the planet-hosting stars. We monitored this field 
from JCBT continuously in I-band  on 05 Feb 2018 at 13:40-16:45 UT and
in V-band on 04 Jan 2019 at 14:10-17:10 UT, on 05 Jan 2019 at 14:00-17:25 UT,
on 06 Jan 2019 at 17:25-19:10 UT and on 08 Jan 2019 at 13:20-18:35 UT. 
This exercise was a part of our search program for new planets. 
We performed differential photometry of each target star in the field 
with respect to the ensemble average of a few other stars in the field to get
the light curves for the target star. The temporal fluctuations in the light curves 
obtained for the stars in that field are not more than $0.3\%$ and none of the light curves
 shows any signature of a transit event of a Jupiter-sized or Neptune-sized close-in planet within the span of our observation. However, we can not comment if any of the light curves could be attributed to a transit event of an Earth or smaller sized close-in planet or a distant (longer period) planet. We report here the light curves of 3 stars in the field namely TYC 3337-1778-1, 
TYC 3337-1676-1 and TYC 3337-83-1. The corresponding light curves, 
after wavelet denoising, are shown in Figure~\ref{fig:ngc1545}. 
The average fluctuations in the baselines w.r.t. the mean flux levels 
and the PSNR values for these stars are presented in Table~\ref{tab:base}.

\begin{deluxetable}{cccccccc}[!ht]
\tablecaption{Parameters quantifying the fluctuations in the light curves with no transit signal\label{tab:base}}
\tabletypesize{\scriptsize}
\tablehead{\\
Date        & Parameters         & WASP-33 (before     & WASP-33 (after      & WASP-50       & TYC 3337-1778-1 & TYC 3337-1676-1 & TYC 3337-83-1 \\
            &                    & GP noise modeling) & GP noise modeling) &               &                                   
            &                 &
}
\startdata
04 Jan 2019 & fluctuation ($\%$) & $-$                 & $-$                 & $-$           & $0.12\pm0.007$  & $0.16\pm0.01$   & $0.16\pm0.01$ \\        
            & PSNR               & $-$                 & $-$                 & $-$           & $865\pm55$      & $639\pm43$      & $714\pm61$    \\
05 Jan 2019 & fluctuation ($\%$) & $-$                 & $-$                 & $-$           & $0.14\pm0.01$   & $0.12\pm0.009$  & $0.16\pm0.01$ \\
            & PSNR               & $-$                 & $-$                 & $-$           & $713\pm63$      & $872\pm71$      & $605\pm40$    \\
06 Jan 2019 & fluctuation ($\%$) & $-$                 & $-$                 & $-$           & $0.15\pm0.01$   & $0.13\pm0.01$   & $0.18\pm0.02$ \\        
            & PSNR               & $-$                 & $-$                 & $-$           & $655\pm70$      & $765\pm74$      & $563\pm52$    \\
08 Jan 2019 & fluctuation ($\%$) & $-$                 & $-$                 & $-$           & $0.16\pm0.009$  & $0.18\pm0.007$  & $0.26\pm0.01$ \\
            & PSNR               & $-$                 & $-$                 & $-$           & $642\pm41$      & $552\pm21$      & $394\pm22$    \\
09 Jan 2019 & fluctuation ($\%$) & $-$                 & $-$                 & $0.1\pm0.007$ & $-$             & $-$             & $-$           \\
            & PSNR               & $-$                 & $-$                 & $1119\pm84$   & $-$             & $-$             & $-$           \\
10 Jan 2019 & fluctuation ($\%$) & $0.5\pm0.05$        & $0.3\pm0.04$        & $-$           & $-$             & $-$             & $-$           \\
            & PSNR               & $202\pm19$          & $297\pm37$          & $-$           & $-$             & $-$             & $-$           \\  
\enddata
\tablecomments{The value of each parameter is shown along with 1-$\sigma$ error margin. Also, some of the values are shown as $-$, which implies that no observation of that star has been made on that day.}
\end{deluxetable}

\clearpage

\section{Results and Discussion} \label{sec:rd}

From the search program for new planets, we have found no indication of any
planetary transit event so far. However, these negative results along with
the observations of the known planet-hosting stars during the out-transit
epochs help us characterize the baseline stability and 
to calculate the lower limit of the transit depth detectable by JCBT and HCT.

\begin{figure}[!ht]
\centering
\includegraphics[scale=0.33,angle=0]{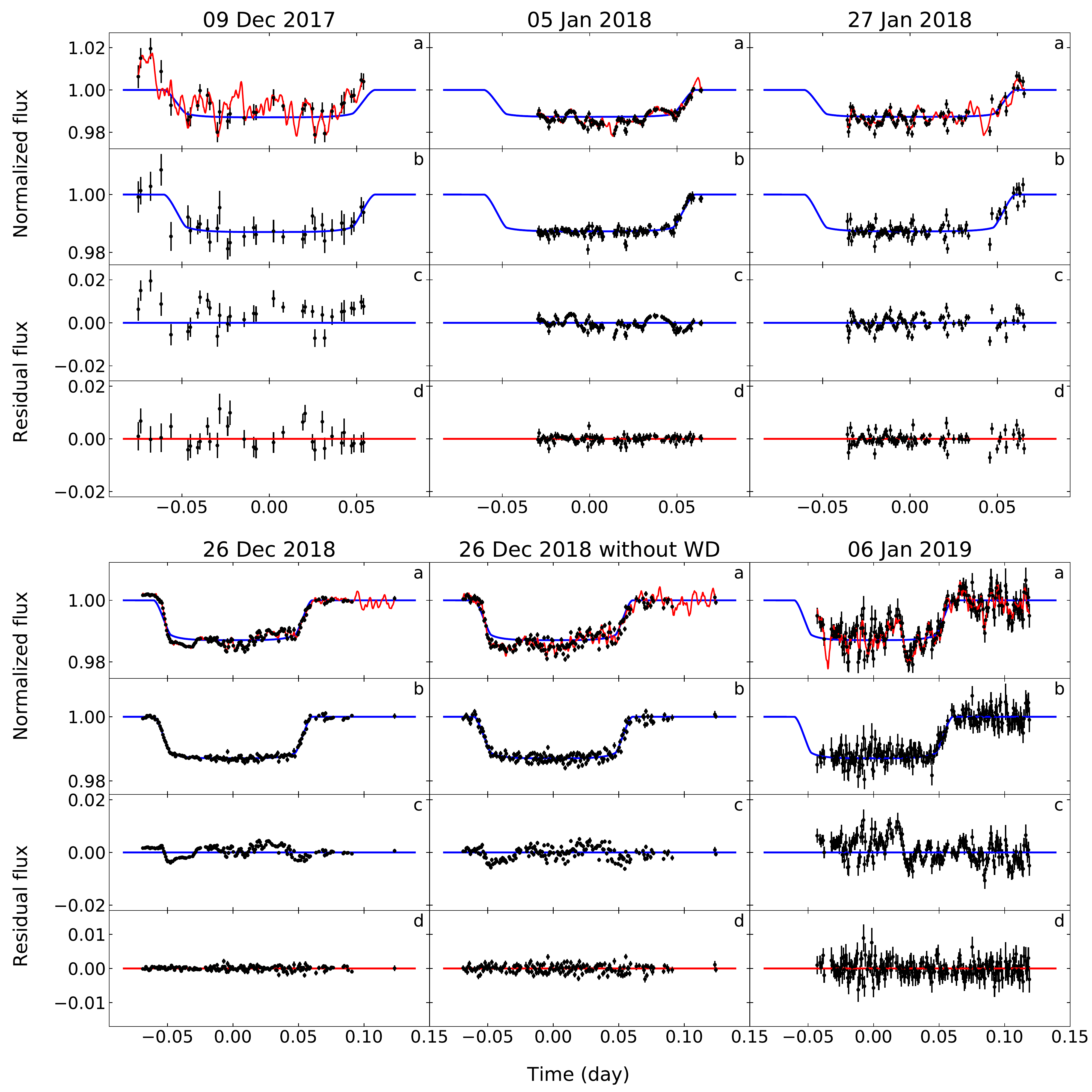}
\caption{The normalized light curves, with and without the wavelet denoising (WD) process and the model fits for WASP-33b. The zero points on the time axes are set at the the mid-transit ephemerides as shown in Table~\ref{tab:obs}.\\
a - The black errorbars represent the normalized wavelet-denoised flux with associated error. On top of it the MCMC-fitted transit models with and without Gaussian process correlated noise (GP) models are shown with red and blue lines respectively. b - The black error-bars represent the normalized wavelet-denoised data minus the GP noise model. On top of it the MCMC-fitted transit models (without GP) are shown in blue lines. c - The black error-bars represent the residual flux with error after subtracting only the transit models (without GP). d - The black errrorbars represent the residual flux plus error after subtracting both transit model and GP noise model.
\label{fig:wasp33bgp}}
\end{figure}

\begin{figure}[!ht]
\centering
\includegraphics[scale=0.33,angle=0]{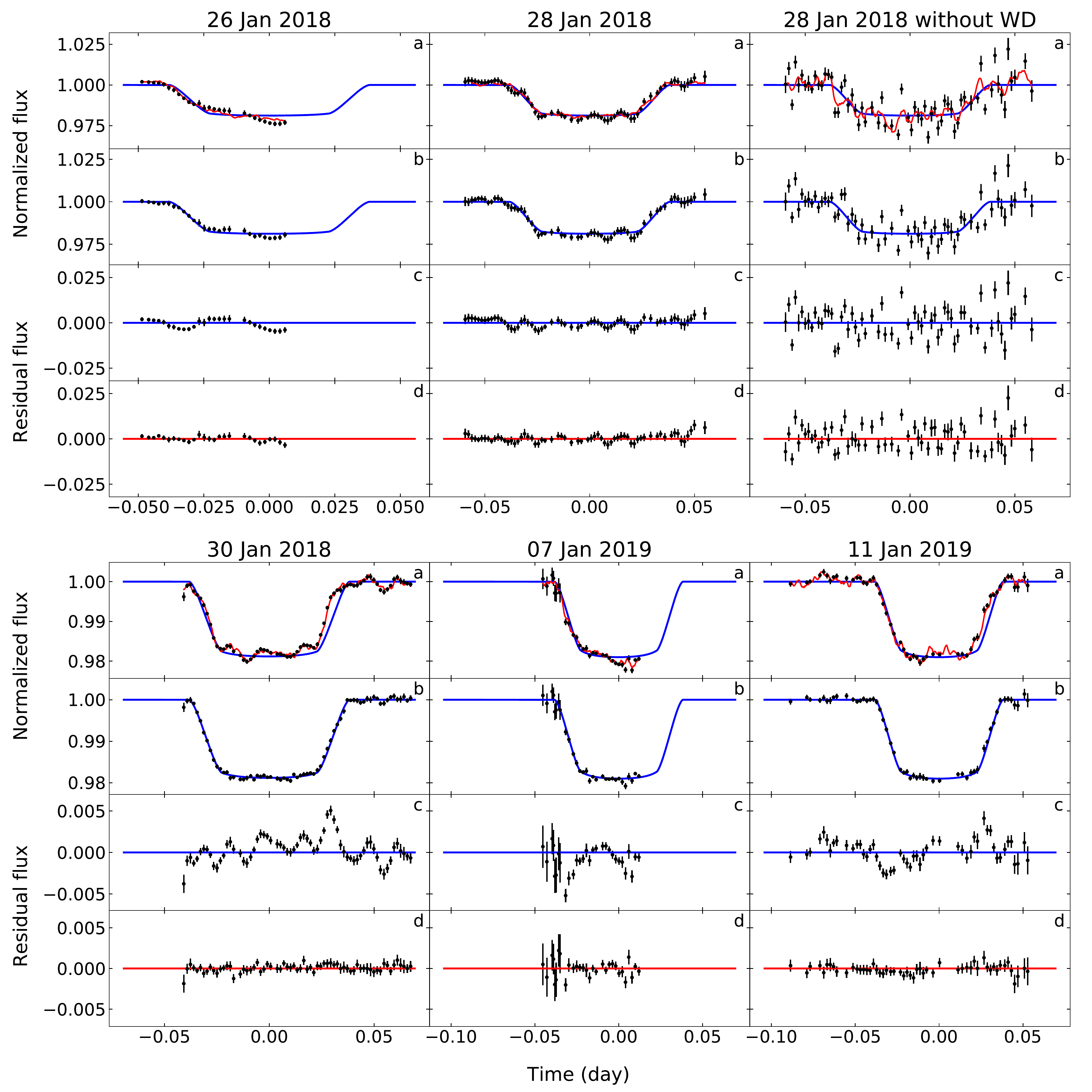}
\caption{The normalized light curves, with and without the wavelet denoising (WD) process and the model fits for WASP-50b. The zero points on the time axes are set at the the mid-transit ephemerides as shown in Table~\ref{tab:obs}.\\
a - The black errorbars represent the normalized wavelet-denoised flux with associated error. On top of it the MCMC-fitted transit models with and without Gaussian process correlated noise (GP) models are shown with red and blue lines respectively. b - The black error-bars represent the normalized wavelet-denoised data minus the GP noise model. On top of it the MCMC-fitted transit models (without GP) are shown in blue lines. c - The black error-bars represent the residual flux with error after subtracting only the transit models (without GP). d - The black errrorbars represent the residual flux plus error after subtracting both transit model and GP noise model.
\label{fig:wasp50bgp}}
\end{figure}

\begin{figure}[!ht]
\centering
\includegraphics[scale=0.33,angle=0]{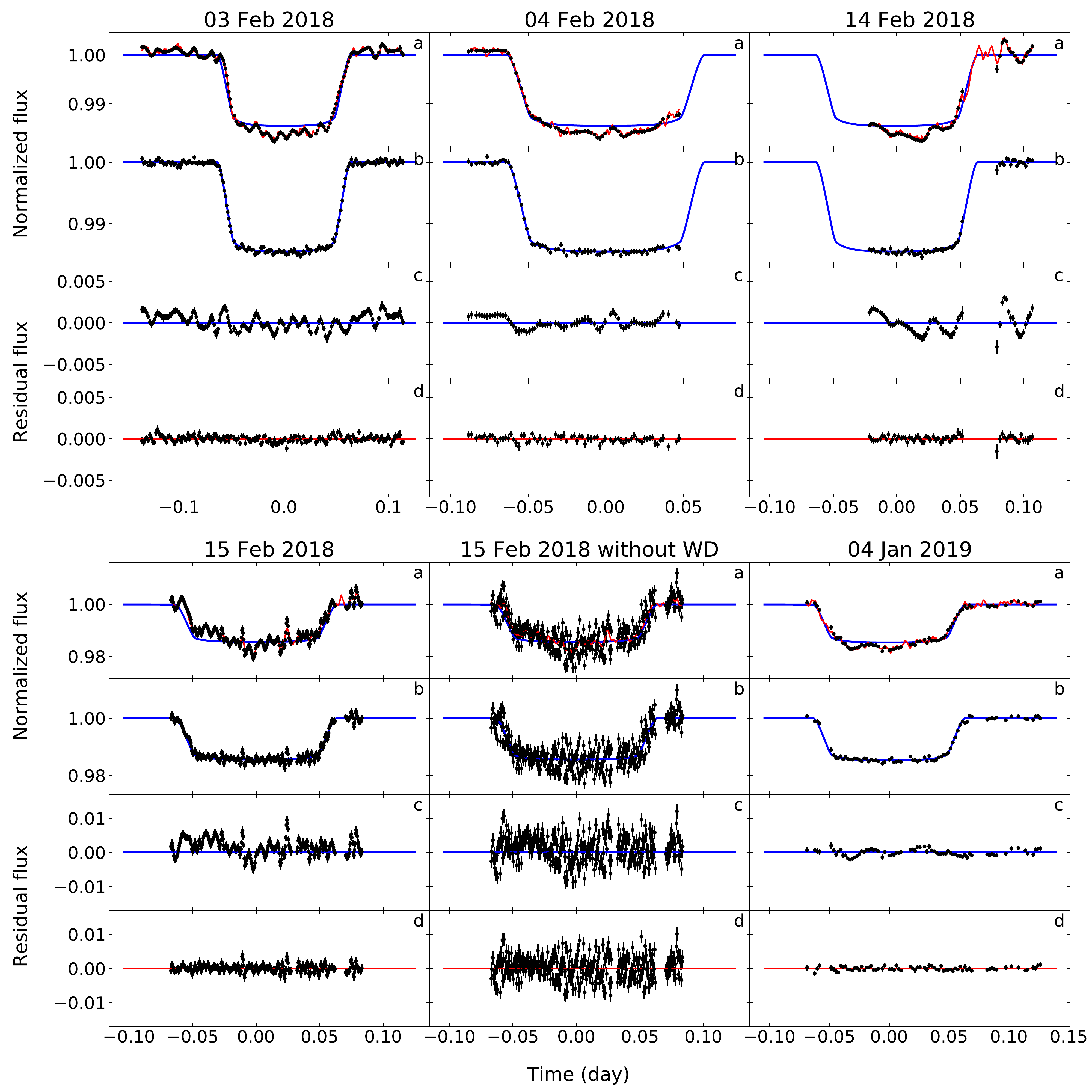}
\caption{The normalized light curves, with and without the wavelet denoising (WD) process and the model fits for WASP-12b. The zero points on the time axes are set at the the mid-transit ephemerides as shown in Table~\ref{tab:obs}.\\
a - The black errorbars represent the normalized wavelet-denoised flux with associated error. On top of it the MCMC-fitted transit models with and without Gaussian process correlated noise (GP) models are shown with red and blue lines respectively. b - The black error-bars represent the normalized wavelet-denoised data minus the GP noise model. On top of it the MCMC-fitted transit models (without GP) are shown in blue lines. c - The black error-bars represent the residual flux with error after subtracting only the transit models (without GP). d - The black errrorbars represent the residual flux plus error after subtracting both transit model and GP noise model.
\label{fig:wasp12bgp}}
\end{figure}

\begin{figure}[!ht]
\centering
\includegraphics[scale=0.33,angle=0]{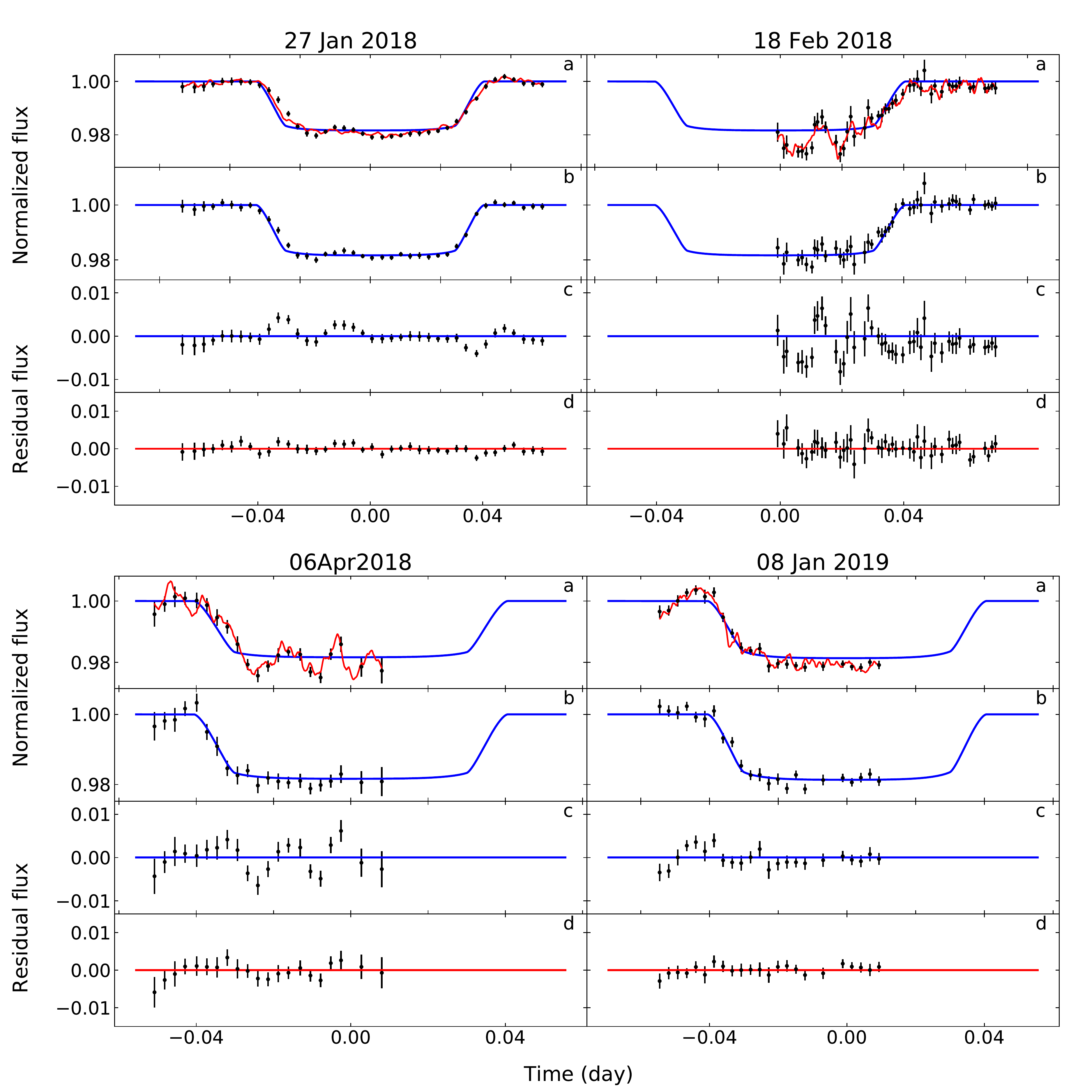}
\caption{The normalized light curves and the model fits for HATS-18b. The zero points on the time axes are set at the the mid-transit ephemerides as shown in Table~\ref{tab:obs}.\\
a - The black errorbars represent the normalized wavelet-denoised flux with associated error. On top of it the MCMC-fitted transit models with and without Gaussian process correlated noise (GP) models are shown with red and blue lines respectively. b - The black error-bars represent the normalized wavelet-denoised data minus the GP noise model. On top of it the MCMC-fitted transit models (without GP) are shown in blue lines. c - The black error-bars represent the residual flux with error after subtracting only the transit models (without GP). d - The black errrorbars represent the residual flux plus error after subtracting both transit model and GP noise model.
\label{fig:hats18bgp}}
\end{figure}

\begin{figure}[!ht]
\centering
\includegraphics[scale=0.33,angle=0]{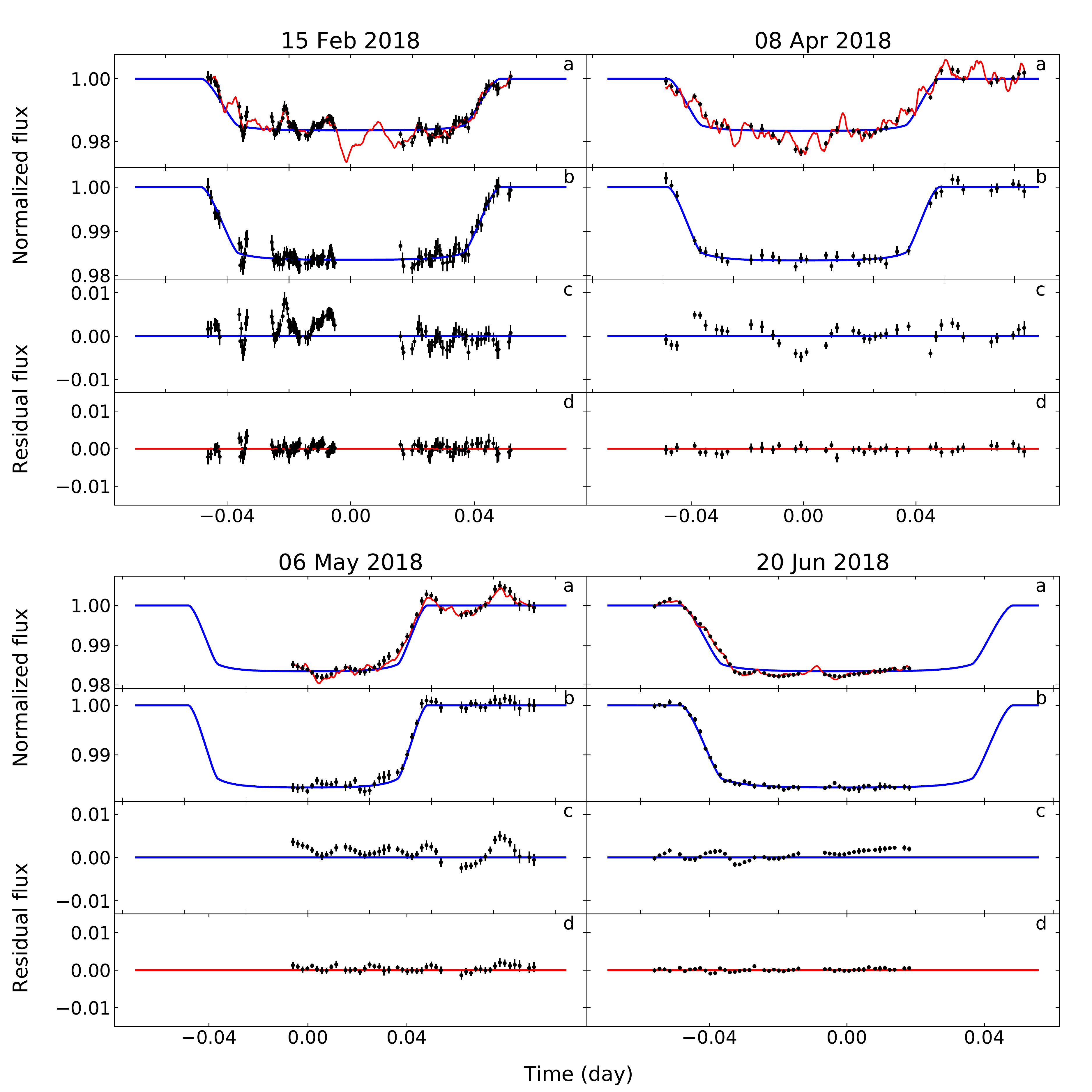}
\caption{The normalized light curves and the model fits for HAT-P-36b. The zero points on the time axes are set at the the mid-transit ephemerides as shown in Table~\ref{tab:obs}.\\
a - The black errorbars represent the normalized wavelet-denoised flux with associated error. On top of it the MCMC-fitted transit models with and without Gaussian process correlated noise (GP) models are shown with red and blue lines respectively. b - The black error-bars represent the normalized wavelet-denoised data minus the GP noise model. On top of it the MCMC-fitted transit models (without GP) are shown in blue lines. c - The black error-bars represent the residual flux with error after subtracting only the transit models (without GP). d - The black errrorbars represent the residual flux plus error after subtracting both transit model and GP noise model.
\label{fig:hatp36bgp}}
\end{figure}

Fitting the observed transit light curves with transit model by using MCMC,
the physical parameters of the hot jupiters are updated with much precise
values. This high precision can be 
attributed to the high photometric S/N and the techniques adopted to reduce
the fluctuations in the light curves. Table~\ref{tab:obs} presents 
 the median of the photometric S/N values of each light curve. The reduced
transit light curves with the best fit models are shown in 
Figure~\ref{fig:wasp33bgp}, Figure~\ref{fig:wasp50bgp}, Figure~\ref{fig:wasp12bgp},
Figure~\ref{fig:hats18bgp} and in Figure~\ref{fig:hatp36bgp}. As evident from 
these figures, the fluctuations in the residual light curves are comparable or
even less to the uncertainties (errors) in the flux values. As we can see
from these figures, after the first stage of preprocessing i.e., WD, the
light curves show different level of fluctuations for different host stars.
Transit light curves for WASP-33b show maximum  fluctuations due to pulsation.
This is consistent with the previous observations by \cite{vonessen14,
johnson15}. These fluctuations could be significantly reduced by GP modeling.
The fluctuations in the transit light curves of WASP-50b, HATS-18b and
HAT-P-36b are found to be moderate after WD and the subsequent GP noise 
modeling has further improved the corresponding light curves. However,
the transit light curves for WASP-12b are found to show minimum  fluctuations
after WD and hence, WD alone would be sufficient for noise reduction in this
case. 

The total transit durations ($T_{14}$) are estimated from the model parameters 
using the relation as follows:
\begin{equation} \label{eq:t14}
 T_{14} = \frac{P}{\pi}\arcsin \left(\frac{\sqrt{(1+R_p/R_*)^2-b^2}}{\sqrt{(a/R_*)^2-b^2}}\right)
\end{equation} 
The masses of the planets were determined by using the radial velocity of the 
host stars and the present updated values of the inclination angle using
the relation:
\begin{equation} \label{eq:mp}
M_p = M_*^{2/3}\left(\frac{P}{2\pi G}\right)^{1/3}\frac{K_{RV}\sqrt{1-e^2}}{\sin i},
\end{equation}
using the fact that, $M_P\ll M_*$. In the absence of any observational 
information, we have assumed circular orbits i.e.,
$e = 0$. For a few targets, the orbital eccentricities have been reported to be 
extremely low \citep{gillon11,turner16} and therefore the assumption of 
circular orbit is justified.

 We derived the surface gravity of the planets $g_p$ as well by using the 
relation \citep{southworth07}:
\begin{equation} \label{eq:loggp}
\log g_p = \log\left(\frac{2\pi K_{RV}\sqrt{1-e^2}(a/R_*)^2}{P\sin i (R_p/R_*)^2}\right)
\end{equation}

We also estimated the equilibrium  temperatures $T_{eq}$ of the planets by 
assuming zero Bond albedo and full re-distribution of the incident stellar
flux. In term of the stellar effective temperature $T_{eff}$, $T_{eq}$ can be
written as    
\begin{equation} \label{eq:teq}
T_{eq} = T_{eff}\left(\frac{R_*}{2a}\right)^{1/2}
\end{equation}
The values of $M_p$, $T_{eq}$ and $\log g_p$ derived from the modeling of the transit light curves preprocessed with wavelet denoising are presented in Table-\ref{tab:par}. The same without the wavelet denoising process are presented in Table-\ref{tab:parnw}.

\vspace{0.1\textwidth}

\clearpage

\section{CONCLUSION} \label{sec:c}

We have observed the transit events of five hot Jupiters of masses ranging
from 1.17 $M_J$ (WASP-33b) to 1.98 $M_J$ (HATS-18b) and radii ranging from 
1.15 $R_J$ (HAT-P-36b) to 1.82 $R_J$ (WASP-12b) by using two facilities in 
India at different places- 1.3m  JCBT and 2m HCT. We have obtained the transit
light curves of these targets with very high transit S/N (transit-depth/noise).
This high transit S/N can be ascribed to the high photometric S/N owing to
the large apertures of the telescopes used. However, apart from  the 
noise emanating from the stellar pulsation, we find that noise from  the 
fluctuating sky transparency contaminates the transit signals significantly.
This is a major drawback of the ground-based observations even with
sufficiently large aperture of the telescope used. In the present work, 
we have demonstrated that wavelet denoising can efficiently suppress the
uncorrelated noises to a great extent. We have also shown that the 
correlated noises can be estimated with  high accuracy and can be
subtracted from the time-series photometric data by using the Gaussian
process correlated noise modeling. 
 Thus, we could update the transit parameters of the planets with very high
precision (less 1-$\sigma$ error) compared to the previously published 
results. Hence, by combining the host star properties, the physical parameters obtained through radial
velocity method and that obtained by precise transit observations, 
the values for the mass, radius, mean density, surface gravity etc.
of the planets are obtained with improved precision.

 Finally, the high stability ($\sim 500 ppm$) of the light curves obtained 
from the observations of the stars during out-transit epoch implies that the
observational facilities as well as the back-end instruments are capable
of detecting the signature of planets not yet discovered.

\section{Acknowledgements}\label{sec:a}
We thank to the supporting staff at the Indian Astronomical Observatory (IAO), Hanle; the Centre For Research and Education in Science and Technology (CREST), Hosakote and the Vainu Bappu Observatory (VBO), Kavalur. We have used PyRAF for most of the tasks of reduction and photometry. PyRAF is a product of the Space Telescope Science Institute, which is operated by AURA for NASA. We thank K. Sankarasubramanian for his insightful inputs and M. Gillon for his help at the initial stage of this work. AC thanks C. S. Stalin, B. Kumar, S. Rakshit and  D. V. S. Phanindra for their helpful suggestions. We thank the reviewer for all the useful comments.\\
\facilities{HCT (HFOSC), JCBT}
\software{barycorr \citep{barycorr}, George \citep{george}, pywt \citep{pywt}, emcee \citep{emcee}}

\clearpage         
 
\bibliography{ms}

\end{document}